\documentclass[sigconf]{acmart}
\AtBeginDocument{%
  }
    
\usepackage{tabularx}

\acmISBN{978-1-4503-XXXX-X/26/02}

\copyrightyear{2026}
\acmYear{2026}
\acmConference[CHI '26]{Proceedings of the 2026 CHI Conference on Human Factors in Computing Systems}{April 13--17, 2026}{Barcelona, Spain}
\acmBooktitle{Proceedings of the 2026 CHI Conference on Human Factors in Computing Systems (CHI '26), April 13--17, 2026, Barcelona, Spain}
\thanks{Accepted to CHI 2026. This is the authors' preprint version.}

\ifdefined\reviewMode

    \newcommand{\rev}[1]{\textcolor{black}{#1}} 

\else

    \newcommand{\rev}[1]{#1} 
\fi

\def\fig#1{Figure~\ref{#1}}

\def\sec#1{Section~\ref{#1}}

\hypersetup{
  pdftitle={Beyond the Desk: Barriers and Future Opportunities for AI to Assist Scientists in Embodied Physical Tasks},
  pdflang={en-US},
  pdfauthor={Hou et al.}
}

\begin{document}
\setcopyright{none}
\title{Beyond the Desk: Barriers and Future Opportunities for AI to Assist Scientists in Embodied Physical Tasks}

\author{Irene Hou}
\affiliation{%
  \institution{UC San Diego}
  \city{La Jolla, CA}
  \country{USA}}
\email{ihou@ucsd.edu}
\orcid{0009-0008-0511-7685}

\author{Alexander Qin}
\affiliation{%
  \institution{UC San Diego}
  \city{La Jolla, CA}
  \country{USA}}
\email{aqin@ucsd.edu}
\orcid{0009-0009-5624-5625}

\author{Lauren Cheng}
\affiliation{%
  \institution{UC San Diego}
  \city{La Jolla, CA}
  \country{USA}}
\email{lacheng@ucsd.edu}
\orcid{0009-0007-6488-7500}

\author{Philip J. Guo}
\affiliation{%
  \institution{UC San Diego}
  \city{La Jolla, CA}
  \country{USA}}
\email{pg@ucsd.edu}
\orcid{0000-0002-4579-5754}

\renewcommand{\shortauthors}{Hou et al.}

\begin{abstract}

More scientists are now using AI, but prior studies have examined only how they use it `at the desk' for computer-based work. However, given that scientific work often happens `beyond the desk' at lab and field sites, we conducted the first study of how \rev{scientific practitioners} use AI for embodied physical tasks. We interviewed 12 \rev{scientific practitioners doing hands-on lab and fieldwork} in domains like nuclear fusion, primate cognition, and biochemistry, and found three barriers to AI adoption in these settings: 1) experimental setups are too high-stakes to risk AI errors, 2) constrained environments make it hard to use AI, and 3) AI cannot match the tacit knowledge of humans. Participants then developed speculative designs for future AI assistants to 1) monitor task status, 2) organize lab-wide knowledge, 3) monitor scientists' health, 4) do field scouting, 5) do hands-on chores. Our findings point toward AI as background infrastructure to support physical work rather than replacing human expertise.

\end{abstract}


\begin{teaserfigure}
  \includegraphics[width=\textwidth]{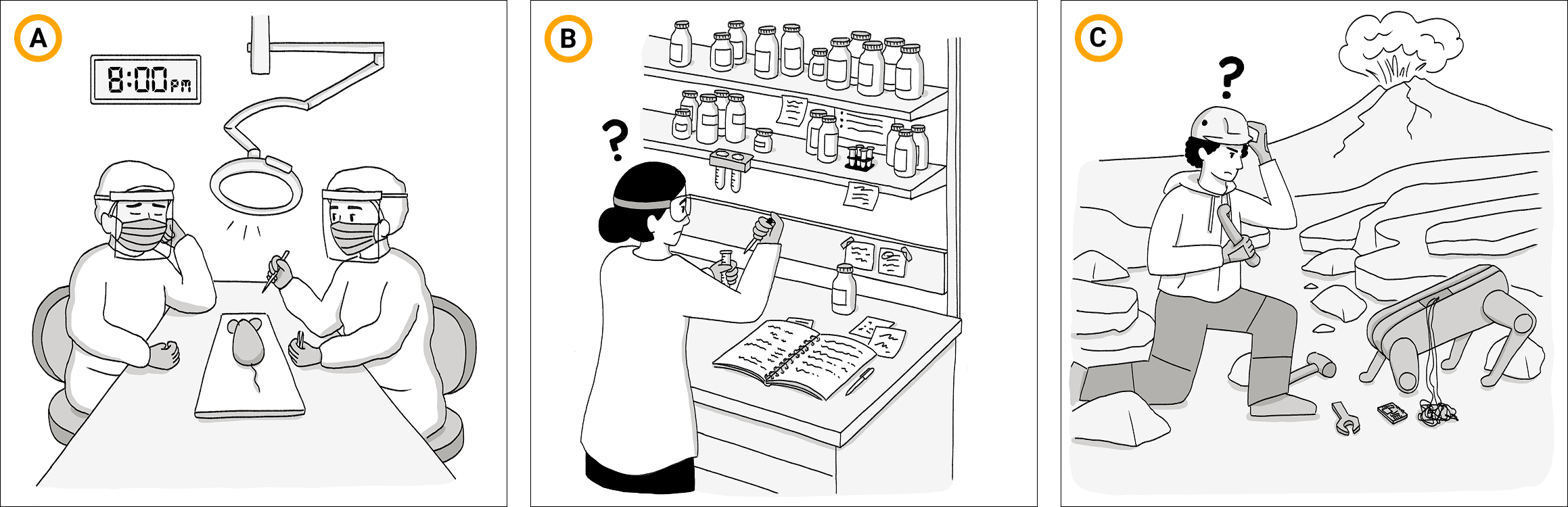}
  \caption{\textnormal{Via interviews with 12 \rev{scientific practitioners working in lab and field settings,} we found three barriers to adopting AI for embodied physical tasks: \textbf{(A)} Neuroscientists like P1 have experimental setups (e.g., surgeries to implant sensors in rodent brains) that are too high-stakes to risk AI making errors. \textbf{(B)} The challenging environments of lab and field sites, like a biochemist's clean-room lab bench with delicate equipment (P12), make it hard to access AI tools. \textbf{(C)} Scientists feel AI cannot match the tacit knowledge and contextual judgment of human experts like a field roboticist improvising materials to get their experimental rig working at a remote volcano site (P2). \textit{(Image credit: all illustrations were drawn by human artist Lauren Cheng without AI assistance.)}}}
  \Description{Three illustrations labeled (A), (B), and (C) depict scientists working in a lab or in a field setting. Image (A) shows two scientists sitting at a table performing rodent brain surgery, with one appearing exhausted as the clock shows 8:00pm. In (B), a cellular biologist stands at a lab bench preparing reagents with a test tube in her left hand and a pipette in her right hand. She appears confused as she relies on scattered sticky notes and analog protocols. In (C), a scientist working at remote terrain near a volcano is stuck fixing an unresponsive robot with limited tools and no support. He holds the broken robot leg in his right hand as the quadrupedal robot lies next to him with tangled wires.}
  \label{fig:teaser}
  \vspace{3pt}
\end{teaserfigure}

\maketitle

\section{Introduction}

\begin{figure*}
    \centering
    \includegraphics[width=\linewidth]{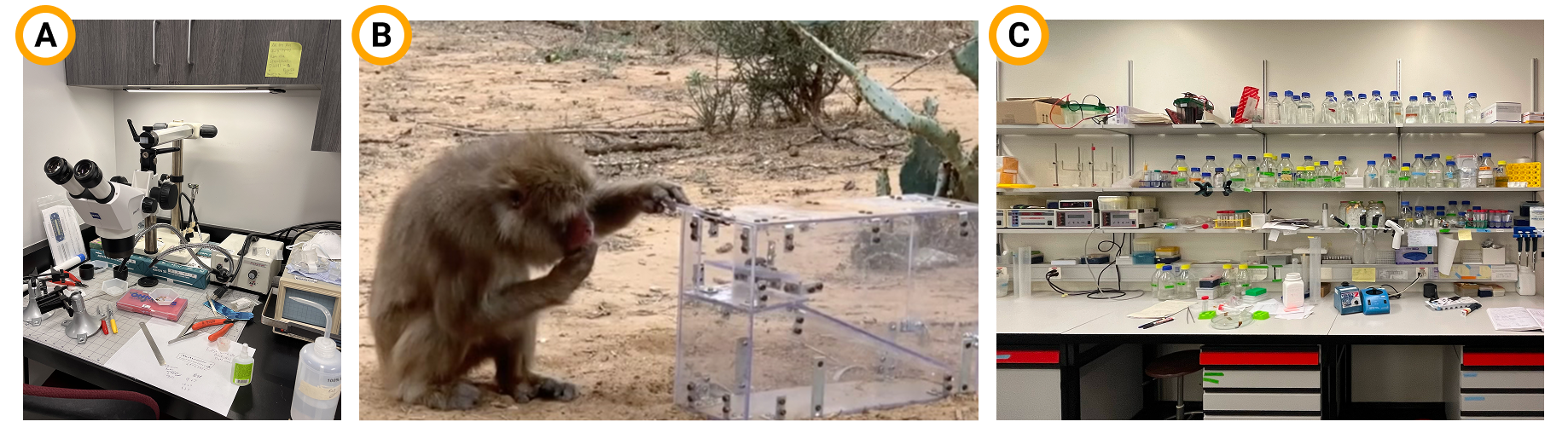}
    \caption{\textnormal{To find barriers and opportunities for AI adoption, we visited workplaces and performed on-site interviews with 12 \rev{scientific practitioners} in a range of lab and field settings such as \textbf{(A)} behavioral neuroscience labs with surgery and bench areas (P1, P5), \textbf{(B)} field site for a primate behavioral scientist (P8), \textbf{(C)} cellular biochemistry lab (P12).}}
    \Description{Four images of various scientific settings are labeled with (A), (B), (C), and (D). Image (A) shows a behavioral neuroscience lab bench with a microscope, tools, and scattered notes for soldering a neurobiology implant device. In (B), multiple scientists in cleanroom suits stand around computers and machinery in a nuclear fusion laboratory. In (C), a macaque in a dry landscape interacts with an apparatus made out of acrylic panels. Image (D) provides a closer look at a lab bench, featuring many reagent bottles and sticky notes on the shelves and a notebook on the table top.}
    \label{fig:science-setting}
\end{figure*}

As AI continues to advance, global investment toward AI for science and the automation of knowledge discovery have led to widely publicized, high-profile breakthroughs. In 2024, teams were awarded the Nobel Prize in Chemistry for AlphaFold, an AI system that uses deep learning to predict 3D protein structures~\cite{jumper2021highly}, along with the Nobel Prize in Physics for the theoretical foundations that led to deep neural networks~\cite{nobelprizephysics2024}. Other fields have seen similar innovations, from drug discovery~\cite{wallach2015atomnet, insilico2024phase2} to climate prediction~\cite{lam2023learning}. Tech companies have also recently invested in using AI for science, such as Google's \emph{AI Co-Scientist} to aid in hypothesis deliberation, literature review, and data synthesis~\cite{gottweis2025towards}.

However, most of these initiatives have centered on computation, simulation, data analysis, and other forms of knowledge work that take place ``at the desk.'' While AI systems are promising for the computational and simulation-heavy areas of science, this is only one aspect of scientific practice. Decades of STS (Science and Technology Studies) and HCI research have shown that discovery frequently involves material interaction~\cite{latour2013laboratory} and embodied physical labor in labs and field sites~\cite{cetina1999epistemic}. A lot of critical scientific labor happens ``beyond the desk'' in high-stakes and physically-demanding environments. For instance, a nuclear physicist must don a full-body cleanroom suit to enter a high-security lab space where they fabricate delicate fuel capsules, a task where fine-grained physical coordination is vital and \emph{a moment's lapse of attention can derail months of hard work and put others in danger}. A biochemist similarly labors under intense timing constraints, running between incubators, centrifuges, and a series of sterile workspaces to keep cell cultures alive while following a choreographed experimental protocol. A field scientist hikes through the forests of the Republic of Congo to study chimpanzee behavior, improvising tools on-the-go in the face of unpredictable terrain and limited infrastructure. These are all real stories from the 12 \rev{\emph{scientific practitioners}}\footnote{\rev{In this paper, we use the term \textbf{\textit{scientific practitioners}} to specifically refer to people engaged in the day-to-day practices that produce scientific knowledge, including experimental setup, data collection, troubleshooting, interpretation, and the embodied labor of laboratory and field environments. Although workers of all ages and experience levels can engage in such practitioner labor, oftentimes much of this labor is taken on by junior lab staff (e.g., graduate students, postdocs, lab technicians) while senior scientific staff (e.g., professors, industry lab PIs) may focus more on ``at the desk'' knowledge work such as writing grant proposals, framing research papers, and giving talks to academic and broader public audiences.}} we interviewed for this paper (see Figures~\ref{fig:teaser} and~\ref{fig:science-setting}), and they illustrate how scientific practice across a range of domains can require physical presence, quick judgment, and expert hand-eye coordination. This sort of \textit{``beyond the desk''} labor currently exceeds what can be codified or simulated on the computer since it involves intimate contact with the physical world.

Despite ``beyond the desk'' science contributing critical findings to research and innovation, little is known about how contemporary AI tools fit into these workflows. For instance, HCI research in AI often focuses on its capabilities for screen-bound knowledge workflows such as programming, data science, creative writing, or UI/UX/visual design~\cite{guo2025from, laban2024beyond,vaithilingam2022expectation, zheng2025towards, liu2023code,shi2023understanding, suh2024luminate}. However, scientific work is often constrained by physical realities and unpredictable environments, involves improvised non-repeatable procedures, and is shaped by the multi-sensory \emph{tacit knowledge}\footnote{\emph{Tacit knowledge} is hands-on, domain-specific knowledge that is hard to precisely articulate in words, and thus rarely written down~\cite{polanyi2009tacit}. In scientific settings this may involve a nuclear physicist knowing how to finely manipulate a fuel capsule `by feel' using their sensory intuition. By definition this \emph{unwritten} knowledge cannot be in the training sets for text-based AI systems like LLMs. One can imagine training AI with video data, but even those cannot capture senses like how some fuel capsules `feel' right or wrong when manipulated with precision handheld tools, and how to adjust on-the-fly.} of domain experts. Moreover, the stakes of error are high, ranging from wasted months of research effort and materials to putting humans in danger. These conditions may pose unique barriers to AI adoption, which led us to the following question that, to our knowledge, we are the first to raise: \textbf{How do \rev{scientific practitioners} who work in embodied, improvisational, and high-stakes lab and field environments perceive the relevance, limitations, and future potential of AI assistance?}

To address this question, we visited lab and field sites of 12 \rev{scientific practitioners} and conducted in-situ interviews in their workplaces. Our participants worked in laboratory and field environments across biology, neuroscience, animal cognition, nuclear physics, materials science, and field robotics. They had diverse backgrounds, from small university research labs to large industrial organizations, and a range of work experience levels \rev{ranging from junior university researchers to senior staff scientists.} Participants reflected on their current use of AI and articulated the stakes and limitations of AI assistance in their domains. At the end of each interview, they each engaged in a speculative design activity where they envisioned an imagined `ideal' future AI assistant.

Our study revealed three barriers that constrain AI adoption in scientific practice, shown in \fig{fig:teaser}: (A) experimental setups are too high-stakes to risk AI errors, (B) constrained environments make it hard to access AI, and (C) AI cannot match the tacit knowledge~\cite{polanyi2009tacit}, contextual judgment, and embodied physical skill of human experts. Thus, instead of seeking AI that ``does the science'' for them, they envisioned future tools that improve human memory, keep track of documentation, and help prevent costly errors due to lapses in human attention. They sketched a series of speculative designs for future AI systems, summarized in \fig{fig:newspecdesign}, that can 1) monitor task status, 2) organize lab-wide knowledge, 3) monitor scientists' health, 4) do field scouting, 5) do hands-on chores. Participants consistently favored passive, context-aware systems that could blend into existing workflows and support the conditions of human-scientific reasoning rather than AI that deprives them of the opportunity to \emph{``do the actual work [myself].''} (P11)

We hope our findings inform ongoing conversations within HCI and the broader Human-AI interaction community about how to design tools that respect the unique constraints of real-world scientific labor. In doing so, we also surface a broader opportunity to \emph{expand the design space of AI beyond raw productivity and automation, and toward supporting the intangible human conditions that make knowledge creation possible.} Our findings can inform AI design for other physically-based domains outside of science, such as medical practitioners, emergency first responders, artisanal craft workers, or field technicians working in demanding outdoor conditions. We make the following contributions to HCI:

\begin{itemize}
  \item The first study examining how \rev{scientific practitioners} perceive AI tools in embodied lab/field work `beyond the desk'
  \item Three sets of barriers that constrain AI use in materially grounded settings such as lab and field science (See \fig{fig:teaser})
  \item Five speculative design concepts reflecting \rev{scientific practitioners}’ future visions for AI support (See \fig{fig:newspecdesign})
\end{itemize}

\section{Related Work}

Our paper extends the long lineage of research on studies of scientists' workflows in two novel ways: 1)~by emphasizing how they envision modern AI helping them, and 2)~by focusing on under-studied physical tasks `beyond the desk.'

\subsection{Studies of Scientists' Workflows}

Scientific workflows and collaborative practices have been long-standing topics of interest in HCI, CSCW, and Science and Technology Studies (STS). For instance, ethnographic studies have examined how scientists conduct their work, from the day-to-day of laboratory life~\cite{latour2013laboratory} to the ``shop work'' and ``shop talk'' through which technical practices, material artifacts, and social processes coordinate to produce scientific knowledge~\cite{lynch1979art}. Viewed through this lens, scientific practice---from the lab to the field---is not only the execution of formal protocols, but also a complex negotiation of material and social conditions. Researchers have documented how tacit knowledge~\cite{collins1974tea}, embodied physical skill~\cite{cetina1999epistemic}, infrastructure~\cite{vertesi2014seamful}, and environmental constraints~\cite{yeh2006butterflynet} shape everything from experimental design to collaboration.

Starting in the 2000s, more contemporary work analyzed scientists' programming and data-centric workflows `at the desk.' Studies have examined how scientists develop and maintain software~\cite{hannah2009how, carver2007software}, rely on computational tools~\cite{huang2025scientists}, manage file dependencies~\cite{gori2020fileweaver}, and organize high performance computing (HPC) scripts, directories, and datasets~\cite{strong2011alamos}.

Although prior work has offered rich accounts of how scientists work to produce scientific knowledge, most of it predates the widespread availability of modern AI, especially its use `beyond the desk' out in lab and field settings. Our interview study advances this lineage of research by \rev{being the first to examine how scientific practitioners engaged in physical lab and field work} imagine modern AI assistance within their day-to-day embodied workflows.

\subsection{Human-Centered AI in Science}

With recent advances in deep learning and generative AI, researchers have explored how these technologies can automate aspects of scientific discovery in tasks like molecular structure prediction, chemical synthesis, and hypothesis generation~\cite{ramos2025review, jumper2021highly, reddy2025towards, baek2024researchagent, rapp2024self}. While pushing the frontiers of what AI can do for science, most efforts center around automating processes of modeling, simulation, and algorithmic optimization~\cite{lu2024ai, jablonka202314} with GPU-enhanced computation.

There is an emerging line of work on robotic arms to automate some physical lab processes like pipetting and diluting microfluids; however, these systems are currently expensive and inflexible, so they are limited more to larger-scale industrial production settings and not seen in most academic labs, where these tasks are done by hand~\cite{robotic_lab_assistants_2024}. \rev{Semi-autonomous UAVs and field robots have considerable potential to support field scientists in tasks such as interacting with remote environments~\cite{outreachrobotics}, terrain imaging~\cite{seifert2019influence}, and environmental monitoring~\cite{koukouvelas2023uav}. Emerging work in robotics for fieldwork has begun to explore opportunities in this direction, highlighting challenges in animal species and individual identification, site access, and data handling~\cite{pringle2025opportunities}. However, existing systems are typically deployed for predefined missions, rather than multi-step or improvisational scientific workflows, and research in this area remains focused on technical and algorithmic capabilities.}

Recognizing the limits of full automation, HCI researchers have focused on Human-AI collaboration, where human scientists drive inquiry and AI assists~\cite{schmidgall2025agent, o2023bioplanner}. These systems support tasks like experimental planning and validation~\cite{schmidgall2025agent, o2023bioplanner}, data processing~\cite{jablonka202314}, and research question ideation (e.g. CoQuest~\cite{liu2024how}, PersonaFlow~\cite{liu2025personaflow}).

Most work here has been on building new system prototypes; to our knowledge, there have been only three prior \emph{studies} of how scientists use AI in their workflows. The first in 2023 presented interviews with professors and tech-industry scientists \rev{(often at more senior levels)} about their desired use cases for AI in science; participants envisioned only AI support for knowledge workflows `at the desk', with potential in science education, data wrangling, literature review, coding, and technical writing~\cite{morris2023scientists}. Two more recent studies at CHI 2025 focused again only on desk-based workflows: one studied how scientists and operations staff use an internal ChatGPT-style chatbot at a U.S. national laboratory (e.g., for writing emails, reports, and manuscripts)~\cite{wagman2025generative}, and the other on how scientists use AI for programming and data analysis~\cite{obrien2025how}.

However, despite the rise of computational techniques, much of modern science still happens in the lab or out in the field, and thus exercise scientists' physical skills and embodied cognitive abilities that go `beyond the desk.' Our work differs from these prior studies by focusing on the perspectives of \rev{scientific practitioners} in improvisational, uncertain, and noisy environments such as wet labs or remote outdoor field sites. \rev{Our study provides a situated, embodied perspective of practitioners ranging from graduate students to experienced full-time lab staff, which complement the findings of prior studies focused on the desk-based computer-centric workflows of scientists~\cite{morris2023scientists,wagman2025generative}.} We extend their findings by situating \rev{scientific practitioners'} perceptions, barriers to adoption, and speculative visions of AI within physical contexts. \rev{Taken together, our findings can be combined with those from prior work on scientists' use of AI at-the-desk to give the field a more holistic view of how lab and field science might be done end-to-end in the future with human-centered AI support.}

\subsection{Human-Centered AI in General Knowledge Work} 

Zooming out farther beyond applications to science, consumer-available AI tools capable of programming, writing, and generating imagery have impacted many types of professional expert stakeholders, especially software engineers~\cite{zheng2025towards, vaithilingam2022expectation, khojah2024beyond, liu2023code}, designers~\cite{subramonyam2025prototyping, zhu2018explainable, shi2023understanding}, educators~\cite{prather2025beyond}, writers~\cite{guo2025from, ippolito2022creative}, artists~\cite{chang2023the, tang2024whats, suh2021ai}, and knowledge workers across areas like law, medicine, and business~\cite{amershi2019guidelines, fok2024marco}. Recent studies have been directed towards understanding how AI could change or improve these domain workflows in terms of productivity, creativity, or collaboration~\cite{vanukuru2025designing, pu2025assistance, suh2024luminate}. Studies have also investigated how professionals use AI as a co-creative or assistive partner, such as surfacing tensions in how writers choose to integrate AI assistance~\cite{guo2025from}.

However, the aforementioned domains and studies typically involve `desk work.' One notable exception~\cite{kernan2023harnessing} examined how LLM-based cognitive assistants can support factory workers in physically-demanding environments, for example, in helping workers resolve mechanical issues with production lines. Similarly, Kernan et al. proposes a set of design guidelines that emphasizes real-time data and domain-specific knowledge integration for AI assistants in manufacturing~\cite{kernan2023harnessing}. While this line of research considers factory-style production labor, our study spotlights AI in the context of the behind-the-scenes physical work that powers science.
\section{Methods}

To understand how scientists are engaging with AI tools in laboratory and field environments, including current use patterns, barriers to adoption, and future speculative design ideas, we conducted situated on-site interviews with 12 \rev{scientific practitioners who work in lab and field settings}. While interviews were planned for 45--60 minutes, many continued longer based on voluntary participant interest. Participants received \$30 USD gift cards, and this protocol was approved by our Institutional Review Board.

\subsection{Participants}

We sought out a range of \rev{scientific practitioners (5 female, 7 male)}, in particular those who worked in settings that required physical presence at the lab bench or in the field (Table~\ref{tab:participants}). Participants were recruited via direct email outreach, word-of-mouth, and snowball sampling. This resulted in 12 participants, spanning R1 university~\cite{wikipedia_r1} labs to industrial research organizations (mostly in the U.S. but a few in the U.K. and Switzerland). Many operated in high-stakes or physically-demanding contexts, such as nuclear science, remote field robotics, applied neuroscience, and ecological fieldwork. Participants \rev{represented a broad range of expertise. Their roles ranged from junior university researchers to senior staff scientists (2 to 28 years of experience in their field).} \rev{In the `Role' column of Table \ref{tab:participants}, the term `Scientist' specifically refers to participants who are formally employed under that job title and often have a doctorate degree. But note in interviews all participants self-identified as `scientists' and were paid practitioners engaged in scientific labor.}

\begin{table*}[h!]
\centering
\caption{We interviewed 12 \rev{scientific practitioners} at research institutions across multiple fields about the barriers and opportunities for AI tools in their daily working environment. The `YoE` column indicates how many years of experience they have in their respective field. `R1 Public/Private' refers to U.S. high research output (R1) universities~\cite{wikipedia_r1}. `AI Use' is self-reported frequency of AI usage.}

\setlength{\tabcolsep}{3.1pt} 
\renewcommand{\arraystretch}{1.5} 
\footnotesize
\begin{tabular*}{\textwidth}{@{\extracolsep{\fill}} l l l l l l l l l}
\toprule
\textbf{ID} & \textbf{YoE} & \textbf{Gender} & \textbf{Field} & \textbf{Institute} & \textbf{Lab Description} & \textbf{Role} & \textbf{AI Use} \\
\midrule
P1 & 3 & M & Behavioral Neuroscience & R1 Public & Neural activity of rodent behavior & PhD student & Sporadic \\
P2 & 5 & M & Field Robotics & R1 Private & Robot locomotion and navigation dynamics & PhD student & Daily \\
P3 & 2 & F & Field Robotics & R1 Private & Robot locomotion and navigation dynamics & Undergrad researcher & Daily \\
P4 & 5 & M & Nuclear Fusion & Industry & Inertial fusion materials design and development & Scientist & Sporadic \\
P5 & 3 & M & Behavioral Neuroscience & R1 Public & Neural activity of rodent behavior & PhD student & Sporadic \\
P6 & 10 & M & Nuclear Fusion & Industry & Inertial fusion materials design and development & Scientist & Sporadic \\
P7 & 3 & F & Nuclear Fusion & Industry & Inertial fusion materials design and development & Data Scientist & Sporadic \\
P8 & 8 & M & Cognitive Science & R1 Public & Social behavior and cognition of primates & PhD student & Daily \\
P9 & 28 & F & Materials Science & Industry & Medical device bioengineering & Staff Scientist & Rarely \\
P10 & 5 & M & Neuroengineering & Industry & Biohybrid neural systems, cognitive preservation & Scientist/Entrepreneur & Daily \\
P11 & 6 & F & Experimental Psychology & R1 Public & Spatial memory and cognition & PhD student & Daily \\
P12 & 4 & F & Biochemistry/Cell Biology & R1 Public & Cellular physiology in cancer and kidney disease & PhD student & Sporadic \\


\end{tabular*}
\label{tab:participants}
\end{table*}

\subsection{Design Rationale for Our Situated Interview Protocol}

Our interview study drew on the method of \emph{naturalistic inquiry}~\cite{lincoln1985naturalistic} to guide data collection situated within participants’ real-world scientific environments. Our protocol consisted of having participants walk through how they typically work on-site (e.g., at their lab bench) and mention where AI plays a role, discuss barriers to adopting AI in their workplace, then come up with speculative designs for how future AI tools could help them. Our protocol includes:

\smallskip \noindent \textbf{1)~Situated on-site interviews}:
We held interviews directly in participants' workspaces when possible, such as at their lab bench. This allowed them to demonstrate workflow components while talking and pointing to surrounding tools, which elicited nuanced qualitative insights.
In the few cases when in-person access was not feasible (e.g., visitors not permitted inside a clean-room), participants were asked to bring photos of their workspace to `virtually' walk through them. This enabled them to reference concrete tasks, tools, and features of their work environment.

\smallskip \noindent \textbf{2) What does `AI' mean? } An important consideration in this study is defining what the term `AI' means. In accordance with our \emph{naturalistic inquiry}~\cite{lincoln1985naturalistic} approach, we started each interview by asking each participant how they use `AI' without explicitly defining the term for them or showing upfront examples. We adopted an \emph{emic} approach (as opposed to \emph{etic})~\cite{emic_etic} by having participants self-define this term based on their own perceptions. In practice, unsurprisingly, most discussed ChatGPT or other chat-based, text-generating LLMs from the 2020s; but some nuclear scientists talked about 2010-era ML algorithms that had been developed in-house for classifying their bespoke datasets. Nobody mentioned GOFAI (i.e., symbolic or logic-based AI that predated the rise of machine learning in the 2000s)~\cite{gofai}.

\smallskip \noindent \textbf{3)~Speculative design activity}: The final part of each interview (15--20 minutes) drew on methods from the field of speculative design~\cite{auger2013speculative, hoffman2022speculative} to elicit system ideas  unconstrained by current technologies or practices. Participants sketched designs for an `ideal' future AI tool to assist their specific workflow.
To help participants think beyond current practices and familiarize them with modern AI tools, the interviewer first asked guiding questions grounded in the workflows each participant described, then demonstrated relevant modern AI capabilities such as multimodal vision and language models they may not have encountered. This enabled them to expand upon their \emph{emic} self-definition~\cite{emic_etic} of `AI' that they initially provided to us. One scenario we presented was how someone building a custom PC can upload photos of a computer motherboard to ChatGPT’s vision model and ask it to identify a LED pin that they needed to use.\footnote{This scenario was intentionally chosen to be simple and domain-agnostic so as not to bias participants toward coming up with any specific kinds of speculative designs in their own field. It was also presented only \emph{after} participants described the challenges inherent to their own physical environment and workflow.}

While this domain-agnostic scenario helped orient participants unfamiliar with multimodal AI, its primary purpose was to encourage participants to think beyond text-based interfaces. Speculative designs emerged mainly through contextual discussion of the barriers previously surfaced and through physical walkthroughs of their lab spaces, which primed participants to think in terms of their own domain. Participants were prompted to sketch their lab spaces while they came up with ideas, using the physical layout to imagine where and how AI could assist their embodied work.

\subsection{Data Analysis}

Interviews were audio-recorded with consent then transcribed by an online service. Two authors reviewed this data independently to familiarize themselves with transcribed content. Transcripts were manually reviewed for errors and analyzed with an inductive approach using thematic analysis~\cite{braun2006using}. Our team iteratively generated and refined codes that reflected emerging patterns across multiple interviews. Codes were grouped by present-day AI use cases, perceptions of current barriers, and speculative design categories. For the speculative design activity, participants' sketches were coded alongside transcript data, which allowed the integration of design ideas into the emerging thematic structure. Throughout the four-month interview and analysis period, the research team met regularly to discuss interpretations and resolve disagreements before finalizing the set of themes presented in Sections~\ref{sec:usecases}-\ref{sec:spec-design}.

\begin{figure*}
    \centering
    \includegraphics[width=0.99\linewidth]{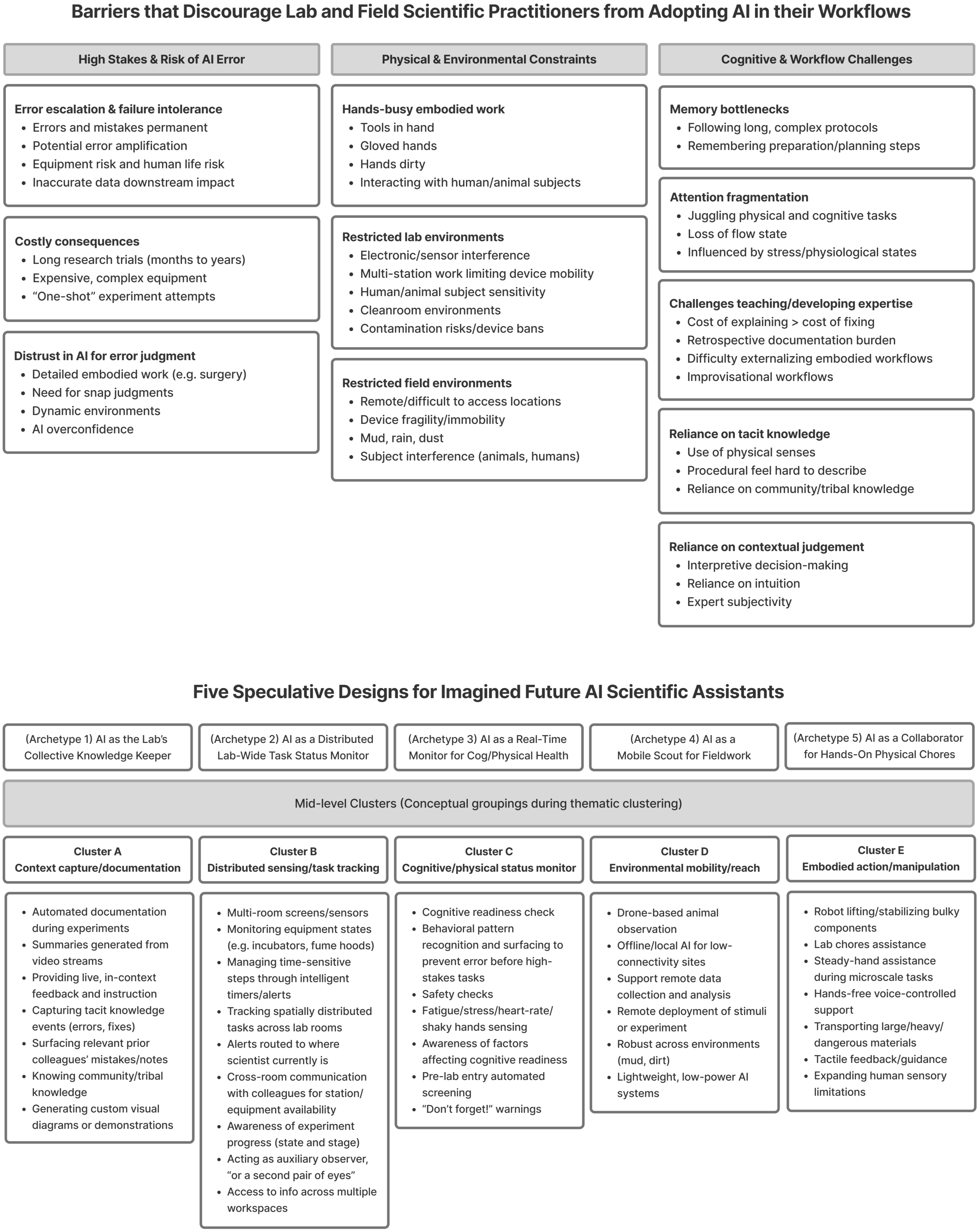}
    \caption{\textnormal{Thematic maps summarizing (top) the three categories of barriers that discourage lab and field scientific practitioners from adopting AI, and (bottom) five speculative AI assistant archetypes organized into mid-level conceptual clusters.}}
    \Description{The figure two thematic maps, one titled ``Barriers that Discourage Lab and Field Scientific Practitioners from Adopting AI in their Workflows," and the other titled "Five Speculative Designs for Imagined Future AI Scientific Assistants.'' Each column represents a major theme and includes the relevant codes within.}
    \label{fig:methods1}
\end{figure*}

Here is a representative example of our iterative analysis process at work: initially, we mapped the raw data of reported use cases and barriers to tasks located in specific lab or field sites. However, the diversity of labs and field sites made this approach too narrow. Thus, we abstracted out to higher-level themes (e.g. `at the desk' versus `beyond the desk') that cut across contexts, which let us capture a wider range of experiences while preserving situational relevance. To analyze speculative designs, we applied a similar procedure. We began by categorizing designs that were tied to discrete physical spaces and types of tasks (e.g. pipetting assistance, surgery protocol assistance). Then, we merged them into broader archetypes, which allowed us to focus on participants' higher-level task rationale. For instance, pipetting assistance would be considered seeking support in `hands-on physical chores.' \rev{See Figure~\ref{fig:methods1} for the thematic mappings that emerged from our analyses. We include two thematic maps to illustrate the structure of analyses across both empirical and speculative components of our study. The top panel summarizes three higher-level barriers that emerged from inductively coding participants’ descriptions of lab and fieldwork. The bottom panel summarizes the five speculative AI assistant archetypes and their mid-level clusters, showing how participants’ design ideas were organized into broader functional groupings.}

~

\subsection{Study Scope and Limitations}

We intentionally scoped our interview protocol and speculative design activity to focus on the day-to-day `on the ground' work of scientific practice. Thus, our study does not cover higher-level considerations such as how organizations set AI policies and cope with systemic risks; see Wagman et al.'s CHI 2025 paper for thoughtful coverage of organizational and social issues around AI adoption at a large U.S. national laboratory (Argonne)~\cite{wagman2025generative}. We also did not cover broader issues such as AI ethics or philosophical objections to AI usage. Note that our participants were likely self-selected to be more open to using AI in their work, so we are lacking the perspectives of those who are strongly opposed to AI use. 

The ethics and norms around AI use in science continues to be debated, especially over challenges related to bias and lack of transparency~\cite{ding2025generative}. Morris previously identified barriers to adoption including scientists' concerns with inaccuracy, hallucination, and falsified results~\cite{morris2023scientists}; other fears include privacy, security vulnerabilities, and plagiarism~\cite{wagman2025generative, nature2025aiwriting}. While some participants mentioned these topics in passing, they were not the focus of our study. 

Although we sampled participants across a range of disciplines (e.g., nuclear science, biochemistry, primate field cognition, neuroscience), we cannot make claims about how universally applicable our findings are across \emph{all} of science. It is likely that we did not cover fields that may be more pro-AI or anti-AI than our sample. \rev{Additionally, while participants represented a range of expertise, our sample skewed towards early-career scientists-in-training and student researchers. Thus, we cannot claim broader generality across \textit{all} scientists (especially professors and lab PIs who may spend more time on grant-writing, advising, and giving talks to academic and broader public audiences). We scoped the focus of our study on \emph{scientific practitioners}--the often-junior staff who do the day-to-day physical labor in lab and field settings. We direct the reader to findings from interviews with professors and industry lab PIs for complementary perspectives from more senior scientists~\cite{wagman2025generative,morris2023scientists}.}

Additionally, all of our participants were based in the U.S. or parts of Western Europe (U.K. and Switzerland). Our findings may not be representative of AI perceptions in geographical regions with different scientific infrastructures, funding models, or cultural norms (See Table~\ref{tab:participants} for participant demographics).

Lastly, while we were allowed to visit and conduct interviews in-situ at lab and field sites, we did not directly observe these scientists doing their tasks live due to the time-sensitive and high-stakes nature of their work (e.g., we could not observe a rodent neural implantation surgery). Instead they walked us through a `simulation' of what they would ordinarily do while they pointed to their physical tools, which limited the fidelity of our observations. And while we showed participants unfamiliar with AI a domain-agnostic multimodal AI example to demonstrate capabilities beyond text-based chatbots, this may have subconsciously anchored some of their speculative designs. Additional illustrative examples may have primed participants to ideate more broadly.

\section{Findings: How Lab and Field \rev{Scientific Practitioners} Currently Use AI Tools}
\label{sec:usecases}

In the following three sections, we first examine how \rev{scientific practitioners working in lab and field settings} currently incorporate AI tools into their work (\sec{sec:usecases}). We then describe the barriers they identified that limit adoption in practice (\sec{sec:barriers}). Finally, we present five archetypal speculative designs for future AI assistants that \rev{participants} came up with, which illustrate how they envision, in the ideal case, AI supporting both their day-to-day and long-term research (\sec{sec:spec-design}). Below we present two categories of current AI use cases: at the desk, and beyond the desk.


\subsection{AI Tools for Knowledge Work `At the Desk'}

Of the 12 \rev{participants} we interviewed, all had heard of AI tools, particularly recent Generative AI (genAI) tools. However, AI usage patterns varied widely, ranging from little-to-no use (P4, P7, P9) to sporadic reliance (P1, P5, P6, P12) to using on a daily basis (P2, P3, P8, P10, P11). The most commonly-mentioned AI tool was ChatGPT, followed by other text-based LLMs such as Claude, Google Gemini, and Perplexity AI-powered search.

For most participants we interviewed, AI use remained peripheral---used intermittently on the computer (i.e., `at their desk') at the beginning or end stages of their workflows, such as during experimental planning, data analysis, or manuscript writing---and was contextually disconnected from their day-to-day physical work in the lab and the field.

The most prevalent AI use cases were for programming and debugging (P1, P2, P3, P5, P6, P8, P10, P12), followed by brainstorming and ideation around experiment protocols (P1, P2, P3, P8, P10, P11). Several participants used AI for clarifying and organizing their thoughts (P3, P10, P11), while others had integrated or considered integrating it into their scholarly communication practices through paper writing assistance (P2, P4, P7, P12), figure generation (P2, P10, P11), and visualization support for presentations (P10). Information-seeking behaviors included using AI for literature review tasks (P1, P8), personal tutoring for learning new concepts (P3, P5, P11), and general knowledge queries (P1, P2, P3, P6, P7, P8, P10, P11, P12). Other use cases included interpreting technical manuals (e.g. setting up trail cameras) (P8), verifying experimental calculations (P5, P12), writing emails, or assisting with paperwork (P7, P11, P12).

When participants walked the interviewer through their work-spaces, many drew clear distinctions between computational and physical spaces. P1, who works in a wet lab studying the neural activity of rodent behavior through \textit{in vivo} electrophysiology, stated that he \textit{``mainly only uses [AI] for data analysis.''} P1 only uses AI tools on his computer at home or at the computer station in an office next to the wet lab. P5, who works in the same field, also expressed that he \textit{``uses [AI] seldomly. I probably don't utilize it as much as I could for my benefit, but it's because it's mainly for coding.''}

Note that both of these \rev{participants} spend significant amounts of time directly interfacing with animals: training, feeding, socializing them, surgically implanting custom-built ``drives'' to measure neural behavior, and running experiments. In these contexts, where research spans years and involves hands-on, dynamic interaction with living systems in-situ, AI tool utility was perceived as limited and more suited to purely computational work back at their desks.

In sum, the majority of \rev{participants}' current AI usage was for computer-based knowledge work at their desks, which confirms findings from the three known prior studies of scientists and AI use~\cite{morris2023scientists,wagman2025generative,obrien2025how}, along with broader studies of other types of knowledge workers~\cite{guo2025from, khojah2024beyond, pu2025assistance, liu2023code}. Where our study goes beyond prior work is uncovering use cases, barriers, and design opportunities for scientific AI use `beyond the desk,' which we present below.

\subsection{AI Tools for Physical Work `Beyond the Desk'}

Several described using AI tools in their lab or field environments, such as voice-activated assistance when their hands were busy (P8), help during DIY (Do-It-Yourself) engineering tasks (P8), and hardware troubleshooting (P2, P3, P11).

For instance, P8, a field researcher who studies primates (e.g., macaques, orangutans, chimpanzees) in remote locations around the world such as rainforests, often needs to engineer experimental setups on-site to conduct his research. These apparatuses cannot be pre-constructed due to their size and are sometimes built with salvaged or improvised materials on-site. P8 recounted a logistically-challenging experience in the Republic of Congo, where he had to construct an experimental apparatus from \textit{``driftwood that [he] found and random objects that [he] uses.''} P8 once asked ChatGPT’s voice assistant mode to help him construct an apparatus made of acrylic, wood, and epoxy to observe primate inhibitory control: \textit{``I can directly ask [ChatGPT] questions, like trivial stuff when I build something, which I'm not great at. I sort of figure stuff out on the go, but if I ask it, `Oh, I need to cut acrylic glass, what tool should I use, what's the right blade?' ... for even stuff like that, I can use it in the field.''} This use case illustrates how AI tools can help \rev{participants} \textit{``figure out tools and solutions''} (P8) and serve as low-friction, just-in-time support for improvisational DIY problem-solving.

In another case, P11, a \rev{doctoral student researcher in the field of experimental psychology} works with virtual reality and game-controller equipment in her experiments and uses ChatGPT to troubleshoot hardware problems. She once nearly canceled a day of scheduled study participants due to issues with a computer graphics card \textit{``but then I gave it to ChatGPT, [and asked] what are all the possible issues for this? What are some of the workarounds, can I replace the cable? Like, going through all the different diagnostics I can do on my end before I take that equipment to a specialist or get outsourced help.''} She later said that she found this form of AI assistance empowering: \textit{``I'm working on a computer vision project right now, which I have zero expertise in, but with ChatGPT, I feel I'm more empowered to do stuff. I don't have to know every single thing.}''

These examples show how \rev{scientific practitioners} need to pick up a wide range of improvised DIY skills on-the-job, often in domains outside their formal academic training. In P8's case, although his expertise was in primate observation, he pointed out that he needed domain knowledge in mechanical engineering and woodwork. For P11, although she \rev{conducts research in the field of cognitive psychology}, in order to run her experiments, she needed to have domain knowledge in computer hardware.

\section{Findings: Barriers that Discourage Lab and Field \rev{Scientific Practitioners} from Adopting AI in their Workflows}
\label{sec:barriers}

Refer to Figure ~\ref{fig:teaser} on Page 1 to see three barriers that hindered participants from using AI for lab and field work.

\subsection{\rev{Scientific practitioners} do not want to face risk of AI errors since physical experimental setups are too high-stakes}

The most frequently-mentioned barrier to AI usage was the trade-off between potential time saved and the devastating cost of AI-induced errors for physical setups that are hard to setup and maintain. This concern spanned labs both large and small, especially those that worked on projects that required substantial investments of time and money.

For example, P1, a \rev{doctoral student} neuroscience researcher, described the extensive, months-long cycle of his lab's experiments: acquiring and training rodents (spans 6 months), hand-building delicate custom electronic ``drives'' designed to measure neural activity (spans 3 months), surgically implanting these tiny drive devices into rodents, constructing bespoke experimental setups, and running behavioral trials to collect data. In particular, pre-experiment surgeries were especially high-stakes and time-consuming: \textit{``If we are hitting one [brain] region in a surgery we've done before, it can usually take eight hours. The first surgery for [labmate's project] took, I want to say closer to 16 hours, because it was multiple [regions]'' (P1).} In this context, P1 feared relying on AI for informational support because even a single mistake could cost him \textit{``the 10 hours that I was in surgery at that point, but also the months that I had spent training this rat---if the surgery goes wrong and the rat dies.''} P1 summed up the risk in one question: \textit{``How much time does [AI] save versus how much time does it cost me if it's wrong?}'' Figure ~\ref{fig:teaser}A illustrates this example.


Participants described similar stakes at larger industrial labs, with an additional element of errors being dangerous and/or amplifying due to AI inaccuracy. P4, P6, and P7's scientific work revolve around designing fuel capsules for experiments at U.S. national labs aiming to achieve nuclear fusion. These experiments rely on rare and costly ``beamtime''~\cite{traweek2009beamtimes} opportunities to  access expensive lab-wide shared laser systems. Collecting sufficient data for even a single experiment can span years, and failed trials cannot easily be repeated (P7). P4 further emphasizes that: \textit{``It's imperative that every data set that we acquire is correct [...] there's a lot more at stake in that measurement. If you're limited to 10 opportunities, if you mess up one, then you can't get that back. Any delays that we incur are going to affect everyone downstream from this as well; in severe cases, that can jeopardize the shot.''}

That is why P4 preferred human judgment and human validation over AI tools: \textit{``Because of the gravity of the data that we're producing, people are much more comfortable trusting another person to do that.''} P6 added that it was hard to trust AI due to its inability to express nuance: \textit{``ChatGPT's writing style makes it sound like it is 100\% correct. Like it'll confidently tell you things that are actually wrong, and that's really dangerous [...] if I'm proposing a hypothesis or giving my opinion, I want to also have the nuance of saying, `Hey, this is what I think, but it might not be entirely true.'''}

\subsection{Physical environments of lab and field sites make it hard to use current AI tools}

Embodied research often takes place in challenging environments that require \rev{participants} to, for example, collect observational data outdoors (P2, P8) or work in highly-controlled settings to protect sensitive experimental conditions (P1, P4, P6), such as in nuclear fusion materials development. Because a substantial portion of their time is spent in these environments, many \rev{participants} could not physically access electronic devices that run AI tools like ChatGPT. 

For instance, during a tour of P7's nuclear fusion lab, she gestured toward a notebook placed outside of lab doors and explained that all external equipment, including phones, laptops, and paper notebooks, were prohibited inside due to contamination risks, which meant no way to access AI assistants inside. Scientists jotted down notes before entering the lab, where full-body clean-room suits are required: hood, hair bonnet, face veil, gloves, coveralls, and boots. P12, who works with cancer cells, must also work in similar clean-room environments (see Figure ~\ref{fig:teaser}B) using laminated paper documentation rather than electronic devices: \textit{``You always have your gloves on with some reagents on it. Maybe some cells are even on it [...] you take [the laminated paper] with you wherever you go, so you can still follow the protocol. At the end of the day, you will spray it, clean it.''} In another instance, P1 noted that he could not have cell phones or laptops in the experiment room because electromagnetic waves, such as Wi-Fi, may interfere with electrode data collection.

In the field AI-enabled devices were avoided not because of contamination risk, but to prevent behavioral interference with the animals being studied such as when brightly-lit devices or voice dictation might distract primates and harm data collection for P8. He remarked that while he did occasionally use his phone to consult ChatGPT, it often was \textit{"messy when you do testing. I'm wearing gloves. I am in a sandy, grimy environment, where I don't want to take out the phone to type. It's definitely a little bit more rugged and reliable to have pen and paper and a clipboard."}

\subsection{AI cannot match the tacit knowledge and contextual judgment of human scientists}

Laboratory and field science involve lots of tacit knowledge consisting of hard-to-capture `unwritten' expertise and embodied decision making~\cite{polanyi2009tacit}, where scientists rely on physical senses such as sight, touch, and even smell to evaluate problems in situ. Many participants were skeptical of AI’s ability to match humans in these skills.

For instance, Figure ~\ref{fig:teaser}C shows how P3 found ChatGPT unable to help troubleshoot mechanical issues with a robot while out at experimental field sites in deserts and volcanos: \textit{``Because it's hard to explain it to the AI in a way that makes sense. Like, I can't really explain how the robot moves; it's difficult to use my words.''}

P10's workplace once debated whether to use AI to automate the placing of fragile organoids beneath a microscope. However, they decided it was more efficient and reliable for a human to do this work, in part because the lab can be a very ``noisy'' environment that makes current AI tools infeasible:

\begin{quote}\textit{``Say you have your pipette and it falls on the ground...it's dirty, but you have something in your hand, so you put your stuff down, you take the pipette up and replace it. It's contaminated. Maybe you should change your gloves. That's so much noise for just one action that happens with something falling on the ground [...] A human brain is actually unmatchable, that biological brain, because we're made exactly for tasks like that [...] We can very easily adjust based on previous knowledge, gut feeling, intuition}.'' (P10)\end{quote}

Moreover, due to the trial-and-error nature of such work, P10 added that some tasks may only be performed once, which makes training a task-specific AI replacement not worthwhile. P10 described this threshold as a \textit{``golden zone''} where it is \textit{``better to let the human do it.''}

A field researcher also shared how his tacit expertise could not be replicated by AI. Over time, P8 has developed the ability to distinguish between chimpanzee subjects by sight: \textit{``For a group of like, 15 chimpanzees, I can tell them apart. Others cannot [...] at some point, you have an intuition.''} Despite attempts to improve computer vision models, AI tools to identify animal subjects \textit{``is still way behind to be good enough for the kind of complex data we have.''}

P12 explained that AI could only assist with piecemeal tasks, such as drafting a protocol for \textit{``specific positions, or like small pieces of experiment planning.''} However, planning an entire experiment end-to-end requires sequencing tasks across multiple physical stations, a process that relies on one's detailed tacit knowledge of their lab's architectural layout accumulated from firsthand experience. As P12 put it, \textit{“First I go into the cell culture... then I do my RNA extraction... then I go to the other room... AI cannot do this [because it] cannot connect the different [lab] stations. It doesn't know the timeline, the context. It cannot plan a whole experiment for me because it doesn't know how an experiment is set up.''}

Lastly, P4 mentioned how evaluating scientific work could be a surprisingly subjective process, which is what separates an expert from a novice and cannot be feasibly replicated by current AI tools. He explained, \textit{``There isn't necessarily a clear right or wrong in terms of data quality [...] is this within the realm of acceptability? Are there negative influences from, let's say, the environment or instrument that would warrant me to have to retake the data set?''} P4 later used the term \textit{``artisan''} to describe the nature of scientific procedures that often evolve over time, which highlights the interpretive, experience-driven aspects of experimental work.
\section{Findings: Five Speculative Designs for Imagined Future AI Scientific Assistants}
\label{sec:spec-design}

\begin{figure*}
    \centering
    \includegraphics[width=1.0\linewidth]{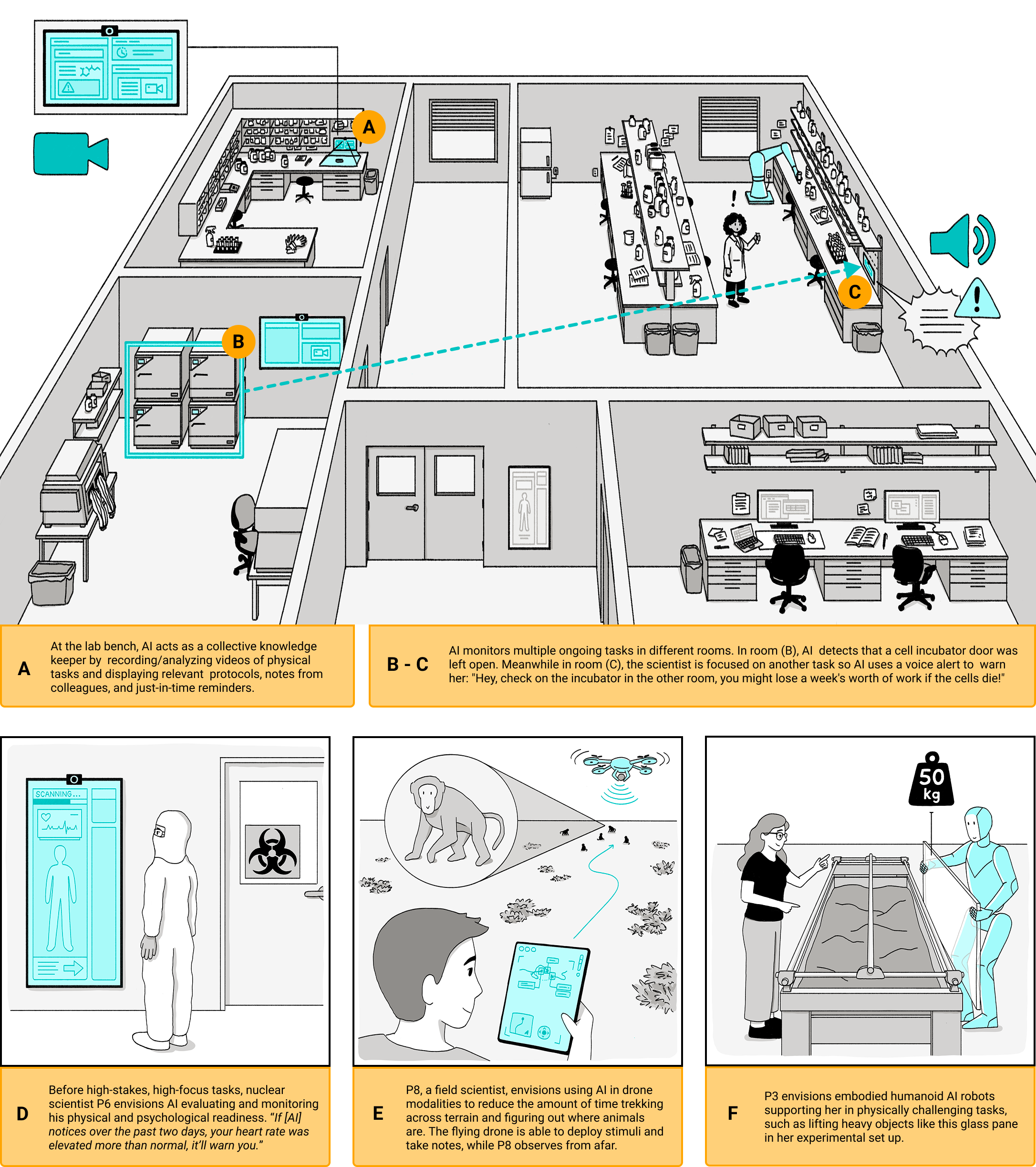}
    \caption{\textnormal{Five archetypes that are representative of speculative designs that our 12 participants envisioned for how future AI could help lab and field scientists: \textbf{(A)} as the lab's collective knowledge keeper, \textbf{(B)-(C)} distributed lab-wide task status monitor, \textbf{(D)} real-time monitor for scientists' cognitive and physical health, \textbf{(E)} mobile scout for fieldwork, \textbf{(F)} collaborator for hands-on physical chores. \textit{(Image credit: all illustrations were drawn by human artist Lauren Cheng without AI assistance.)}}}
    \Description{A larger illustration shows an overhead view of a lab space with a distributed AI system across multiple rooms labeled (A), (B), and (C), while three illustrations underneath labeled (D), (E), and (F) demonstrate scientists using AI to assist them in their work. In the chemical reagent room (A), a screen located on the lab bench wall houses relevant protocols, notes from colleagues, reminders, images, videos, observations, and warnings specific to the lab. A camera on the device is directed at the work surface. In room (B), cell incubators are highlighted to signify detection of a drop in temperature. In another room (C), a scientist focused on a separate task is alerted about the cell incubators by the screen at her lab bench and by an audio warning. Illustration (D) depicts a scientist wearing a cleanroom suit facing a wall-mounted screen displaying “scanning…,” health information, and suggested tasks. Next to them is a door with a biohazard symbol. In (E), a field scientist working in a remote area deploys stimuli to far away primates via a drone, monitoring their behavior on his device. Illustration (F) shows a woman directing a humanoid robot to pick up a fifty kilogram glass panel used in a muddy terrain tank in her robotics lab.}
    \label{fig:newspecdesign}
\end{figure*}

As each participant was conveying barriers to AI adoption in their workplace, that provided a natural segue into our speculative design activity, where we had them sketch out what an `ideal' future AI assistant would look like for typical tasks. They drew a map of their lab or field space on paper and pointed out how an AI might help. Their design ideas varied depending on the type of task and its location within the lab or out on the field (e.g. lab bench, fume hood, incubation room, surgery room, experiment clean-rooms, outdoors, see Figure ~\ref{fig:science-setting} for examples).

Due to space limitations here, we will not present every single design idea that participants had. Instead we grouped similar-themed ideas together into a set of five archetypal speculative designs and present them here.

\subsection{\textbf{Archetype 1: AI as the lab's collective knowledge keeper}} Across nearly all participants, human limitations in knowledge transfer and documentation emerged as persistent bottlenecks. \rev{Participants} spoke about limitations in their ability to recall and pass down expertise (P1, P2, P3, P4, P5, P6, P7, P9, P10, P12). P3 summed it up as: \textit{``No one likes to do documentation when you're actually working. You just want to work [...] but afterwards you don't remember certain parts. It'd be super helpful if Generative AI could do it for you.''} Experienced scientists struggled to pass down contextual knowledge and expertise to junior colleagues due to fragmented documentation and overwhelming context (P1, P2, P4, P6, P9, P12); and junior scientists struggled to navigate lab-specific, idiosyncratic, tacit practices without clear guidance when a mentor was not nearby (P3, P5, P4).

Some labs maintained shelves of note-filled binders that often went unread, while others had abandoned multiple attempts to create usable computer documentation systems or digital archives. Even when documentation did exist, it was frequently described as dense, confusing, and constantly evolving (P1, P3, P4, P7), exacerbated by each lab and organization having their own custom protocols, tools, equipment: \emph{``There's a lot of tribal knowledge''} (P4).

Thus, in our design sessions participants envisioned context-aware AI systems that could alleviate these gaps by capturing, summarizing, and situating knowledge in real time. For example, P2 imagined a mounted AI camera system that could monitor his robotics experiments through video, digest the streamed footage, and \textit{“provide corrections and live feedback in-context''} for the experiment. He envisioned being able to query the AI, which would then refer back to specific video frames when generating responses.

P1 imagined scenarios where future AI could actively participate in data collection and notetaking: \textit{``There's a ton of digital information that it could automatically take the notes from, `Hey, this thing co-occurred with that thing' [...] or integrating behavioral tracking in my note taking like, `At what point does the rat do this?' Or if I just set up a boundary, `Here's the area I want to keep a track of and here's the rat. Make a note when the rat leaves this area.''}

P12 combined a similar idea with lab-based tacit knowledge-sharing, where context was automatically captured and passed on to colleagues: \textit{``Because we have a couple of people in the lab, if AI could note that I made a mistake there, and for everyone else that has to use this protocol again later, tell them `Please be aware this happened to me. That's how you solve it.' Then, [AI] could pop up and tell you, `Someone had a problem here at this step.'''} See Figure ~\ref{fig:newspecdesign} (A).

\subsection{\textbf{Archetype 2: AI as a distributed lab-wide task status monitor}} Whether in the lab or out in the field, participants frequently identified cognitive limitations that frustrated them, such as difficulty recalling specific procedural knowledge (P1, P2, P3, P4, P5, P10, P11, P12) and lapses in memory or attention in the midst of doing hands-on tasks (P1, P4, P5, P8, P11). These moments of cognitive load were seen as opportunities for AI intervention as a persistent memory aid, context-aware feedback provider, or auxiliary observer.

For example, P1 faced attentional limitations during live experiments with rodents since he needed to juggle multiple simultaneous processes: evaluating data (audio and visual), attending to electrophysiology data, and managing other active tasks, such as \textit{``presenting a certain stimulus at a specific time, taking the rats out, cleaning the arena, setting it back up, putting them back in, baiting a maze [...]''} He also needed to simultaneously attend to equipment failures or apparatus malfunctions, then react quickly. P1 imagined that a camera-based AI assistant would be able to offload some of this burden, acting as an auxiliary monitor---an extra pair of eyes---that can provide attentional scaffolding when cognitive load exceeds human capacity: \textit{``I could totally see it being useful in tracking the things that I might normally be looking at but that I just can't during this time because I'm too focused on this other part or something going wrong.''}

P12 designed for similar attentional limitations when multi-tasking in a cell biology wet-lab with many ongoing tasks and sensitive equipment. Her idea in Figure ~\ref{fig:newspecdesign} (B-C) illustrates experimental protocols distributed across screens in different rooms, and AI monitoring multiple ongoing tasks for her. P12 explained how this system could prevent error by surfacing contextually important information for her at the right time, \textit{``Imagine you walk around and you have a screen there like, `Ah, I'm at this step. I need this and this at this station.' Then, at every screen, every station, you follow the next steps.''} P12 added that this could help prevent careless mistakes; she shared a story where a labmate accidentally left a cell incubator door open, which ended the experiment and delayed their work for weeks. A system could detect the open incubator door in the equipment room (where people usually are not present), illustrated in Figure ~\ref{fig:newspecdesign} (B), and then send an audible alert to a scientist working in a different lab room, as shown in Figure ~\ref{fig:newspecdesign} (C).

P5's hands-free voice-interaction design was informed by his workspace, where access to input devices like computer keyboards was not possible: \textit{``My biggest concern in this space is my hands are often full in some way.}'' During the design session P5 demonstrated how he constructs a rodent brain implant drive under time pressure to check on other tasks:

\begin{quote}
    \textit{``Like, I'm holding this [tetrode wire that is thinner than a human hair], and I finally have the position I want, but I'm thinking, `Hey, is this actually gonna fit?' or I'm not sure how much I have on my alarm left before I need to check on the rats. If I could ask AI on the spot without me having to stop and go to a computer, like a distributed system throughout the lab.''}
\end{quote}

The voice-activated AI monitoring system he envisioned (similar in form to Figure ~\ref{fig:newspecdesign} (C)) aligned with his vision of being able to attend to multiple aspects of the lab concurrently while reducing the friction of task-switching.

\subsection{\textbf{Archetype 3: AI as a real-time monitor for scientists' cognitive and physical health}} Besides monitoring lab tasks, participants also envisioned AI systems that act as status monitors \emph{for the actual scientists} who must do high-stakes and high-intensity work. They brought up concerns that dangerous memory and attention lapses stemmed not only from stressful workloads, but from cognitive and physiological fluctuations in their bodies.

In P4's lab, he and his colleagues were encouraged to self-evaluate one's own headspace and physical condition before certain kinds of intensive work with nuclear materials. Drinking too much caffeine the night before could be reason for pause. P4 added, \textit{``You don't want to be in autopilot [...] I make sure not to do this [task] unless I feel like I'm focused.''} P6, who shares this concern, speculatively designed an AI assistant that could screen for these physical and cognitive fluctuations and alert him accordingly before entering the experiment room: \textit{``If it notices over the past two days, your heart rate was elevated more than normal, it'll warn you.''} See Figure ~\ref{fig:newspecdesign} (D) for an illustration of this idea.

More generally, several \rev{participants} (P1, P4, P5, P10, P12) worried that lapses in memory or careless mistakes could lead to unwanted harm and cascading consequences: \textit{``If your work rhythm gets thrown off, you can forget things [...] if you forget to arm [activate] that thing, you don't collect any data. That's common''} (P5). P5 envisioned querying AI before running an experiment as a quick cognitive status check, \textit{``Hey, I'm about to start, am I forgetting anything?''}

\subsection{\textbf{Archetype 4: AI as a mobile scout for fieldwork}} Participants working in the field envisioned AI assistants that were mobile, lightweight, compatible with remote or rugged environments (P2, P3, P8), and powered by local AI models with offline access (P2, P3). Unlike lab-based researchers, these participants often moved between locations and conducted work that required environmental adaptability.

Because voice interactions with AI could disrupt naturalistic conditions or alter subject behavior (e.g., for animals being observed in close proximity---P8 and P11), they explored alternative modalities in speculative designs. P8, who often collects data at primate sanctuaries, imagined using small autonomous flying drones to reduce the amount of time spent trekking across long distances: \textit{``Oftentimes, there is time wasted to figure out where even the animals are.''} P8 speculated on future capabilities of these drones: \textit{``You can probably efficiently collect a lot of information about where animals are and what they're doing by using a drone.''} Beyond passive observation, P8 envisioned active intervention, such as deploying stimuli or food via drone and recording the resulting behaviors: \textit{``Why couldn't the drone do that for me and just also then film it and see what happens?''} These designs, shown in Figure ~\ref{fig:newspecdesign} (E), illustrate how highly-mobile and non-verbal AI modalities align more with the constraints of field-based research.

\subsection{\textbf{Archetype 5: AI as a collaborator for hands-on physical chores}} Many participants struggled with technical chores that were not core to their research questions but still essential to running their experiments: soldering (P1, P5, P3), epoxying (P8), sawing (P8), wiring (P2, P3, P11), cementing (P1, P5), mixing chemical substances (P1, P6, P9, P12), performing surgical procedures (P4, P6, P10). These skills were often self-taught or learned through trial and error, since they had no formal professional training in those fields. In some cases, recently-acquired skills were only useful for a single experiment. As P8 noted, investing significant time in mastering such skills could slow research momentum, moreso when it can be slow to get feedback on them: \textit{``My PI [principal investigator] would get a chance to look at my [hardware devices] a week after I built them and say, `Okay, this one looks really good. This one isn't sturdy enough. This one isn't wound tight enough' [...] but I'd forgotten what I did already.''}

After surfacing these pain points, participants imagined multimodal AI assistants that could provide scaffolded, live support during such embodied hands-on work. Both P1 and P12 suggested modalities that allowed them to visually learn while they picked up new protocols: \textit{``If I could have some audio thing saying, `What's the next step?' Or have a screen that displays the next step instruction or visual instruction [explaining] what this part is [...]'' (P1).} They shared how audio interaction with AI could provide the best hands-free assistance, alongside visual instructions or diagrams on the screen to serve as persistent references, offering clarity during unfamiliar or error-prone procedures.

Human sensory and physical limitations were also surfaced in the interviews. P12 noted that, \textit{``We have things you cannot see by eye [...] you have to train your muscle memory.''} Multiple participants reported challenges related to the steadiness of their hands (P1, P3, P5, P12) or sustained attention and focus (P1, P4, P12), compounding the difficulty of performing hand-eye coordination tasks that required precision. These participants tended to extend their visions into more embodied, robotic forms of AI assistance; their imagined robots were designed to compensate for human sensory or physical limitations like acting as an extra pair of eyes and hands to help them in these chores.

For example, P10 and P3 envisioned humanoid robots that could help lift, transport, or stabilize materials, whether delicate tiny components like neural cells or heavy equipment like glass panels during experiments (Figure~\ref{fig:newspecdesign} (F)). For P3, this also made certain parts of experiments more accessible: \textit{``An extra pair of hands would help most when you have to take glass panels out [...] it's really heavy.''}

\smallskip \noindent \textbf{\rev{Scientific practitioners} preferred passive AI, with proactive intervention for critical errors.} One recurring theme we observed across design sessions was that most participants (P1, P3, P4, P6, P7, P8, P9, P12) leaned toward passive rather than proactive assistance. Passive assistance stays in the background, offering support only when directly queried by the user. In contrast, proactive assistance engages with users by issuing real-time alerts or suggestions based on human behavior. Some participants were more strongly averse to proactive assistance than others. P9 felt it would be disruptive to her workflow and wanted AI systems to notify users through subtle cues: \textit{``I would prefer the robot to raise its hand or flash until I touch it. I don't want it to interact and directly talk to me like, `Hey, you missed that!' That would drive me nuts.''}

Modalities of AI intervention that came up most frequently in sessions were wall-mounted screens and hands-free, voice-interactive systems. Notably, wall-mounted screens \emph{without audio or voice controls} were a preferred modality for participants who were concerned that audio could be a potential confound or distraction when running experiments (P1, P11, P12). In these scenarios, participants preferred AI assistants to provide suggestions, instructions, and warnings via text on large mounted screens, or intervene through minimal visual cues, like flashing red lights (P9). More generally, P11 did not want future AI systems to deprive her of the opportunity to \textit{``do the actual work [myself]}.''

That said, some (P1, P2, P5, P10, P11, P12) favored a mix of passive and proactive. They emphasized that although passive systems were preferred, they wanted AI to intervene when a mistake was imminent, especially in tasks where mistakes could be costly or hard to detect (P2, P12). P5 wanted control over this threshold of AI intervention: \textit{``Ideally, there'd be different preferences. For one, because you obviously don't want [AI] to every 30 seconds go, `Hey, do this. Hey, do that.' But in emergency situations, like, `Hey, you forgot to arm a system,' I want that.''}
\section{Discussion}

Our participants judged AI less by raw capability than by how well it fit the embodied, high-stakes conditions of their work. They found the most value in systems that eased cognitive load through memory, attention, and documentation support, but preferred these to remain passive in the background unless errors became critical. More broadly, this suggests a shift in framing; rather than designing AI as a collaborator that directly participates in scientific reasoning, \emph{AI may be more effective when treated as infrastructure that supports the conditions under which scientific reasoning happens.}

\subsection{Limitations of AI for Physical and Embodied Scientific Work}

Despite rapid advances in recent tools, AI still remains peripheral among our participants in the physically-oriented work of lab and field science. Corroborating the findings of prior studies~\cite{wagman2025generative, morris2023scientists}, it is used mostly `at the desk' for planning, literature review, data analysis, or writing. Despite the status quo, participants were able to use our speculative design activity to envision deeper, context-aware AI integration into their future workflows. This gap highlights structural mismatches, not necessarily technical limitations. As an analogy, current AI assistants resemble `generalist' programmers or technicians who can write boilerplate code but lack the bespoke domain expertise to identify which approaches are the most insightful within a scientific field. Current AI tools are also ill-suited for improvisational and materially grounded practices that characterize much of day-to-day scientific work `beyond the desk.'

While extensive prior work has posited that knowledge-based professions are more readily augmented by, or even displaced by, AI~\cite{tomlinson2025working, brynjolfsson2025canaries, maslej2025artificial, sherman2025amazon, de2023generative}, our study findings suggest that scientific domains involving physical manipulation, situated judgment, and tacit expertise currently resist such displacement of human labor. Tasks requiring bodily coordination and tool improvisation (e.g., meticulous pipetting techniques, DIY repairs in the field) exemplify the kinds of uncertain dynamic settings where scientists viewed human adaptability as irreplaceable. Iterative experimental work often requires scientists to learn new skills or obtain new materials and equipment. Automation needs therefore vary widely depending on individual situations and cannot be easily routinized with AI.

Our findings complement recent analyses on AI's impact on various occupations~\cite{tomlinson2025working}, which shows that roles involving manual labor, operating machinery, or other physically-oriented tasks currently have the least potential for AI collaboration. While Tomlinson et al.~\cite{tomlinson2025working} show that AI aligns best with knowledge work and communication-heavy tasks, our findings reveal how even within knowledge-intensive scientific roles, AI adoption is uneven, concentrated to desk-based activities, and largely absent from lab and fieldwork practices. Although we focused on \rev{scientific practitioners} working in physical domains, we view their work as a proxy for other occupations with similar core elements such as, say, hospital surgeons or emergency first responders.

\subsection{Future Opportunities for AI in Scientific Practice}

Speculative design activities at the end of each interview revealed directions for how AI could be integrated in ways that augment scientists' capabilities without displacing their judgment (\fig{fig:newspecdesign}). From our findings, we derive four higher-level design opportunities for AI, which address recurrent challenges that our participants faced. \rev{These opportunities build directly on the five speculative archetypes described in Section~\ref{sec:spec-design}.}
\\

\textbf{(1) Design to scaffold memory, attention, documentation, and knowledge transfer in physical tasks.} \rev{Archetype 1 (AI as a collective knowledge keeper) and Archetype 2 (AI as a distributed task monitor) show how participants struggle} with fragmented documentation and lapses in attention during complex embodied work, as well as difficulty capturing and understanding tacit knowledge. These \rev{archetypes} point toward designs of real-time, context-aware recording and summarization tools to reduce cognitive load `in the moment,' preserve expertise, and support knowledge transfer. \rev{Unlike desk-based tools that require users to stop and externalize knowledge, support should operate in the background during embodied activity. This distinction matters because physical workflows make it difficult to shift attention or focus away from the task at hand; knowledge must be captured and organized in situ rather than through retrospective, at-desk effort.}
\\

\textbf{(2) Tune AI proactivity to relative stakes.} \rev{Archetype 2 and Archetype 3 (AI as a monitor for scientists' cognitive/physical health) illustrate how participants determine interruptions by AI to be acceptable when errors or safety risks carry meaningful consequences, such as open incubator doors or unsafe physical or cognitive conditions.} \rev{Scientific practitioners} otherwise preferred systems that stayed passive, intervening only when errors had high stakes. Thus, enable AI to adapt its intervention level to task context, with user-configurable thresholds (e.g., use only visual cues in settings that are sensitive to sound, such as with animal subjects).
\\

\textbf{(3) Expand interaction modalities beyond the desk.} Arche-type 2, Archetype 4 (AI as a mobile scout for fieldwork), and Arche-type 5 (AI as a collaborator for physical chores) illustrate that scientists often work with their hands occupied, across multiple rooms and workspaces, or in remote outdoor settings where desktop/tablet interfaces are impractical. Participants envisioned solutions ranging from voice interfaces, large ambient displays, and physical indicators, to flying drones and modular robotic assistants. AI systems need to be embedded in and responsive to the environmental and spatial realities of lab and field work.
\\

\textbf{(4) Embrace digital naturalism in designs.} \rev{Archetype 4 highlights the need for AI systems that can adapt to local environments and are able to operate under mobility and limited infrastructure. Many future} AI designs could benefit from elements of \emph{digital naturalism}, a design philosophy that advocates for modular, field-adaptable, and open-ended scientific tools~\cite{makery2019quitmeyer}. Quitmeyer cautions against stripping away the rich, local contexts of scientific practice and environments~\cite{quitmeyer2024growing}. These tools should be materially embedded and co-configured on site, becoming extensions of scientists’ senses and mental models, and helping them form deeper connections to their work instead of abstracting it away ``at the desk.''

Taken together, these opportunities suggest designing AI as infrastructure integrated into scientists' sites of use.

\subsection{AI as Infrastructure instead of Collaborator: Supporting the conditions for human scientific reasoning}


Current public conversation around AI in science has focused on using it to automate reasoning and discovery, as exemplified by highly-funded corporate projects such as Google DeepMind's AlphaFold, which won the 2024 Nobel Prize for Chemistry~\cite{jumper2021highly}, along with follow-up systems like \emph{AI Co-scientist}, ``a multi-agent AI system [...] to accelerate the clock speed of scientific and biomedical discoveries''~\cite{gottweis2025towards}. Despite the attention given to these large-scale automation efforts, our study participants described the day-to-day of lab and field science as surprisingly ``artisan'' (in the words of P4), requiring hands-on skill, tacit knowledge, and situated improvisation while doing physical chores in the lab.

We saw that many challenges our participants faced were less about enhancing reasoning and \emph{more about sustaining the complex, tightly-coupled physical conditions that make deep scientific reasoning possible}. Participants derived fulfillment from discovery; it was the mental and procedural burdens of minimizing mistakes within their physical environments that they struggled with. Participants consequently did not envision AI as a collaborator in reasoning or decision-making. They instead imagined intelligent, context-sensitive \textbf{\textit{infrastructure}}--- tools that could passively monitor workflows, intervene to prevent errors, and reduce time wasted on unproductive troubleshooting attempts. 

The high stakes of error were surfaced as a major constraint to AI adoption. Mistakes meant weeks of lost labor or, in extreme cases, catastrophic consequences. With these conditions in mind, participants were open to AI systems that prevented lapses in attention or memory, but were skeptical towards AI tools that produced confident outputs without accountability or transparency. While pattern recognition and workflow monitoring were welcome, interpretive AI systems were seen as brittle, lacking the embodied, tacit knowledge that scientific judgment relies on.

These findings suggest a reframing of AI integration in scientific contexts. While much of recent work has focused on building AI systems that can reason, hypothesize, or automate decision-making, our study participants had different unmet needs. Instead, they imagined intelligent tools that strengthen the underlying infrastructures that make scientific reasoning possible: monitoring experiments, stepping in rapidly to prevent human lapses, which includes bodily limitations in memory, attention, and focus, and preserving context across their local communities, spaces, and time. This reframing aligns with prior critiques in human-centered computing, such as concerns that current computational systems risk imposing solutions that prioritize abstraction and \emph{detachment from physical context}---``reinventing virtually every other site of practice in its own image''~\cite{agre1997computation}. Suchman and Agre argue that all research practices, "even the most
analytic, [are] fundamentally concrete and embodied"~\cite{suchman1987plans}, and our findings reflect this. The \rev{scientific practitioners} we interviewed did not see reasoning as something that could be separated from the physical and practical conditions of their work, and were skeptical of AI tools that treated it that way. They advocated for more ``defensive'' speculative AI designs---systems that mitigate risk and preserve stability instead of maximizing speed and efficiency.

Despite advances in technology, many \rev{scientific practitioners} still rely on paper-based tools (e.g., P12’s laminated paper protocols, P8’s field notebooks), underscoring a persistent misalignment between current automation and scientists’ real-world needs. Two decades ago, Yeh et al. observed similar preferences among field biologists and created ButterflyNet, a hybrid physical–digital system to bridge this gap~\cite{yeh2006butterflynet}. Today, with AI and technology far surpassing 2006 capabilities, this is an opportune moment to revisit hybrid system designs grounded in the realities of scientific practice. 

In addition, the growing flexibility and declining costs of modern AI systems make them feasible for ad-hoc appropriation. This is especially relevant in the scrappy, resource-limited environments our participants described. In such settings, bespoke digital tools are costly, niche, and time-consuming to build, risking obsolescence as research directions shift. Fortunately, today’s AI models are becoming less expensive and easier to adapt, offering potential for AI tools that can be rapidly customized by end-users. While laboratory and field science provided the frame for our work, these findings might also offer insights for other high-stakes, embodied, and situated domains, including medical practitioners, artisanal craft workers, and field technicians that work in challenging physical settings.

\section{Conclusion}

This paper is the first, to our knowledge, to report empirical accounts of how laboratory and field scientific practitioners in embodied and improvisational contexts envision the role of AI. Through on-site interviews and speculative design sessions at 12 scientific practitioners' workplaces, we surfaced three barriers to adoption: experimental setups being too high-stakes to risk AI errors, constrained physical environments, and AI's lack of embodied tacit knowledge. Thus, today AI is mostly used for desk-based science work, but our participants were able to sketch ideas for future tools that scaffold memory and attention during hands-on practice and extend across modalities from robots to lab-wide systems. These speculative designs highlight the importance of tacit knowledge~\cite{polanyi2009tacit}, physical constraints, and spatially distributed workflows. In sum, we aim to shift the focus of AI for science from automating reasoning and computational work at the desk to infrastructure that sustains the embodied physical conditions upon which creative reasoning and discovery depend.

\begin{acks}
This work was supported by the Alfred P. Sloan Foundation grant number G-2024-22587. We are grateful to the scientists and scientists-in-training who shared their time, trust, and space with us. 
\end{acks}
\bibliographystyle{ACM-Reference-Format}
\bibliography{sample-base}

\end{document}